\documentclass[twoside,fleqn]{article}
\usepackage{graphicx}
\usepackage{amsmath}
\usepackage{espcrc2}

\title{Deconfining transition in two-flavor QCD}

\author{
    J.M. Carmona\address[ZARA]{Departamento de F\'{\i}sica Te\'orica, 
    Universidad de Zaragoza, 50007 Zaragoza, Spain },
    M. D'Elia\address[GENO]{Dipartimento di Fisica dell'Universit{\`a} di
    Genova and INFN, I-16146, Genova, Italy},
    L. Del Debbio\address[PISA]{Dipartimento di Fisica dell'Universit\`a di Pisa
    and INFN, Via Buonarroti 2 Ed. C, I-56127 Pisa, Italy},
    A. Di Giacomo\addressmark[PISA], 
    B. Lucini\address[OXFO]{Theoretical Physics, University of Oxford,
    1 Keble Road, OX1 3NP Oxford, UK},
    G. Paffuti\addressmark[PISA], C. Pica\addressmark[PISA] \thanks{Speaker at the 
    Conference (Email: pica@df.unipi.it)}} 

\begin{document}

\begin{abstract}
The order and the nature of the finite-temperature phase transition of QCD
with two flavors of dynamical quarks is investigated. An analysis
of the critical exponent of the specific heat is performed through 
finite-size and finite-mass scaling of various susceptibilities. Dual
superconductivity of QCD vacuum is investigated using a disorder
parameter, namely the v.e.v. of a monopole creation operator. Hybrid R
simulations were run at lattice spatial sizes of $12^3$, $16^3$, $20^3$
and $32^3$ and temporal size $N_t=4$, with quark masses in the range
$am_q = 0.3 - 0.01$.
\end{abstract}

\maketitle

\section{Introduction}
\label{sec:introduction}
In QCD with two dynamical fermions a finite temperature phase transition is expected at \mbox{$m_q=0$}. The basic question of the order of this phase transition has been extensively studied in the continuum by an effective sigma model analysis\cite{piwi} and on the lattice with Kogut-Susskind\cite{fuku,colombia,karsch,jlqcd,bernard} and Wilson\cite{cp-pacs} fermions. Nonetheless a conclusive answer has not yet been found.

At zero quark mass in the Pisarski-Wilczek scenario, if the axial anomaly is effectively restored the transition is predicted to be a fluctuation-driven first order transtion; in the opposite case the transition can be second order in $O(4)$ universality class; mean field critical behavior with logarithmic corrections is also possible as a limiting case between the previous two. Since first order phase transitions are generally robust against small perturbations, we expect the first order behavior also if the effective coupling of the anomalous current is small enough. K-S fermions break explicity the $O(4)$ simmetry to $O(2)$ at finite lattice spacing, so in the case of a second order phase transition we expect $O(2)$ critical behavior until the continuum limit is recovered.

The mass term in the QCD Lagrangian corresponds to an external magnetic field in the effective sigma model, thus we expect the transition to remain first order at finite and small quark masses if the transition were first order at $m_q=0$. In the other cases, the mass term completely washes out the phase transition to an analytical crossover. 

To determine which of the above possibilities is actually realized, the standard test of universality is a comparison of the critical exponents. Critical exponents can be extracted from the scaling analysis of susceptibilities obtained from the second derivatives of the partition function:
\begin{eqnarray}
\rho &=& -\frac{\partial}{\partial \beta} \ln \left<\mu \right> = \; \langle\tilde S_g \rangle_{\tilde S_g} - \langle S_g\rangle_{S_g} \\
C_V &=& \frac{1}{VT^2} \frac{\partial^2}{\partial \beta^2} \ln Z \longrightarrow \chi_{ij} , \chi_{ee} , \chi_{ie} \\
\chi_m &=& \frac{T}{V} \frac{\partial^2}{\partial m_q^2} \ln Z = V [\left<(\bar\psi\psi)^2\right>-\left<\bar\psi\psi\right>^2]\\
\chi_{ij} &=& V [\left<P_iP_j\right>-\left<P_i\right>\left<P_j\right>], \quad i,j = \sigma , \tau\\
\chi_{ee} &=& V [\left<(\bar\psi D_0 \psi)^2\right>-\left<\bar\psi D_0 \psi\right>^2] \\
\chi_{ie} &=& V [\left<P_i (\bar\psi D_0 \psi)\right>-\left<P_i\right>\left<\bar\psi D_0 \psi\right>]
\end{eqnarray}
with $V=L^3 N_t$, $D_0$ is temporal component of the Dirac operator and $P_\sigma$, $P_\tau$ indicate the average spacial and temporal plaquette respectively. 
$C_V$ denotes the specific heat and it can be expressed in term of $\chi_{ij}$, $\chi_{ee}$ and $\chi_{ie}$; $\chi_m$ is the susceptibility of the chiral condensate and only its disconnected component is taken into account in the present work.
The first of these susceptibilities $\rho$ is defined in term of the disorder parameter introduced in \cite{I,III}. $\mu$ creates a magnetic charge defined by some given abelian projection and $\left<\mu\right>$ detects magnetic monopoles condensation. The parameter has been tested in quenched theory\cite{I,III} and it is well defined in full QCD\cite{IV}. The independence from the chosen abelian projection has also been successfully tested\cite{III,adriano}.
\begin{figure}[t!]
\includegraphics*[width=\columnwidth]{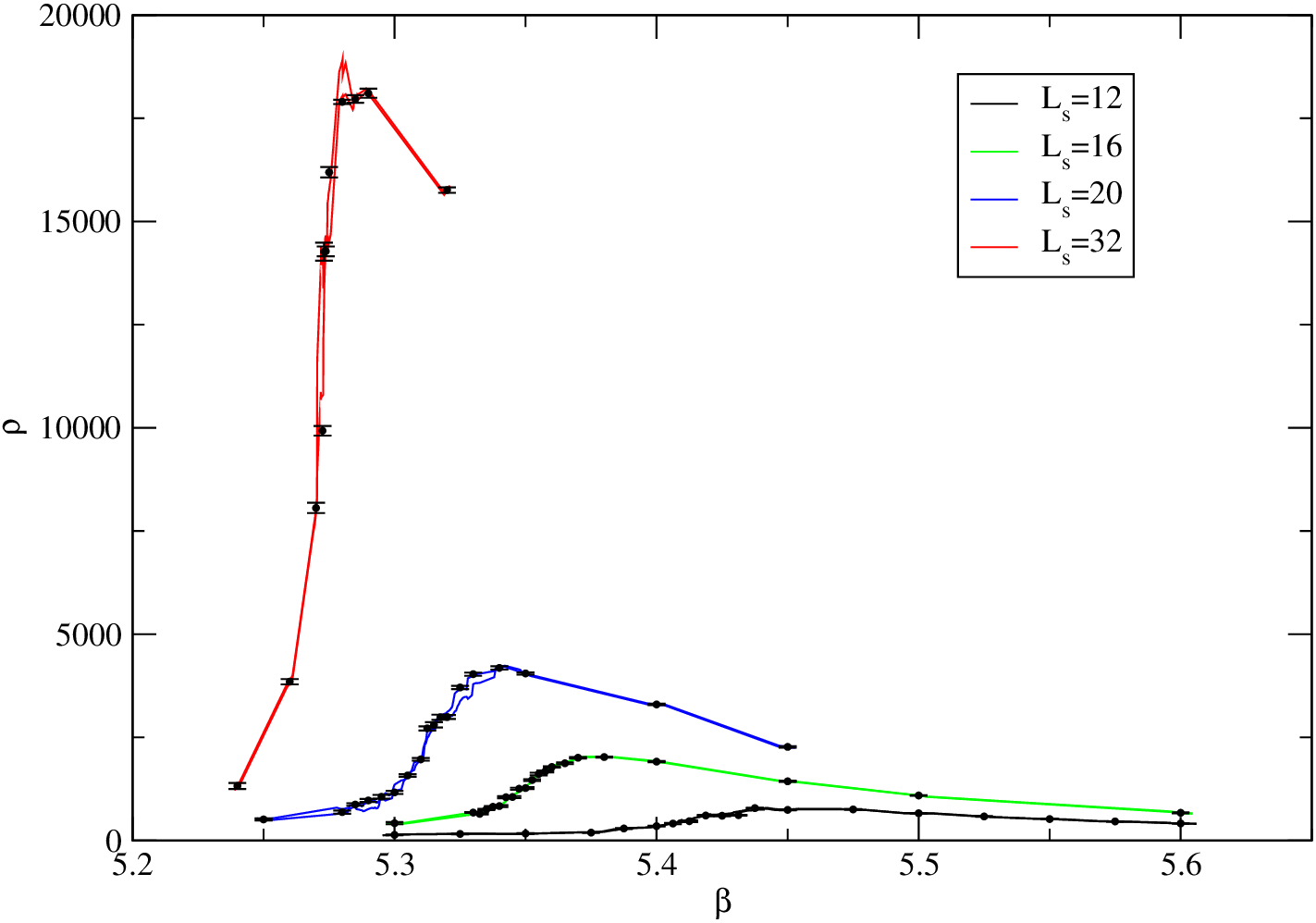}
\vspace{-35pt}
\caption{$\rho$ as a function of L for Run1. Reweighting curves are also shown together with MC data.}\label{rho}
\vspace{-20pt}
\end{figure}
\begin{table}[b!]
\vspace{-20pt}
\caption{Run parameters for the numerical simulations. \mbox{$L_s \cdot m_\pi$} varies in the range $[8.9-15.8]$}\label{runpar}
\begin{tabular}{|c|c|c|c|c|}
\hline & \multicolumn{2}{|c|}{$am_q$} & \multicolumn{2}{|c|}{\# Traj.} \\
\hline $L_s$ & Run1 & Run2 & Run1 & Run2 \\
\hline 12 & 0.153518 & 0.307036 & 22500 & 25000  \\
\hline 16 & 0.075 & 0.15 & 38000 & 54700  \\
\hline 20 & 0.04303 & 0.08606 & 12800 & 12500 \\
\hline 32 & 0.01335 & 0.0267 & 7000 & 15100 \\
\hline
\end{tabular}
\vspace{-20pt}
\end{table}

The peaks of these susceptibilities define the pseudocritical coupling and their heights obey scaling laws. Since we have two parameters, namely the bare quark mass $am_q$ and the finite spatial size $L$, in order to obtain scaling relations of just one variable, we constrained our parameters by the relation $am_q\cdot L^{y_h}=cost$. In proximity of the critical point, the singular part of the free energy density $f_s$ satisfies
\begin{equation}
f_s(t, m_q, L^{-1}) = b^{-d} f_s(b^{y_t}t, b^{y_h}m_q, bL^{-1})
\end{equation}
where $t=(T-T_c)/T_c$ is the reduced temperature, $y_t$ and $y_h$ are the thermal and magnetic critical exponents, and $b$ is an arbitrary scale factor. From this and using the above constraint, we obtain the scaling laws
\begin{eqnarray}
\beta_c(L^{-1}) &=& \beta_c(0) + c_\beta L^{y_t} \label{bcsca}\\
\rho^s(t, L) &=& L^{y_t} \Psi(t L^{y_t}) \label{rhosca}\\
C_V^{max} &=& c_0 + c_1 L^{2y_t-d} \label{cvsca}\\
\chi_m^s(t, L) &=& L^{2y_h-d} \Phi(t L^{y_h}) \; . \label{chisca}
\end{eqnarray}

\section{Numerical Results}
\label{sec:numres}
We used the plaquette action for gluons with the Kogut-Susskind quark action. Configuration updating was performed adopting the standard Hybrid R algorithm. The lattice temporal size was fixed at $N_t=4$, the molecular dynamics step size was $\Delta\tau=0.005$ and the molecular dynamics trajectories were generated with unit lenght. As discussed in the previous section, we constrained simulation parameters by $am_q\cdot L^{y_h}=cost$. Expecting $O(4)$ scaling from previous works, we fixed the magnetic critical exponent to be $y_h=2.49$; notice that the $O(2)$ value is also very near to that. We performed two parallel set of runs, referred to as Run1 and Run2, the only difference between the two beeing that for Run1 we set $am_q\cdot L^{y_h}=74.7$ whereas for Run2 $am_q\cdot L^{y_h}=149.4$. Table\,\ref{runpar} summarizes the simulations parameters for our runs and shows the statistics collected from each run. Results from Run1 suffer from the low statistics collected at $L=32$ and should be considered as preliminary.

\begin{table*}[ht!]\small
\caption{}\label{exp}
\vspace{-10pt}
\begin{tabular}{|c|c|c|c|c|c|c|c|} \hline
\multicolumn{2}{|c|}{$y_t$ $(\rho)$} & \multicolumn{2}{|c|}{$y_t$ $(\chi_{\sigma\sigma})$} & \multicolumn{2}{|c|}{$y_t$ $(\beta_c)$} & \multicolumn{2}{|c|}{$y_h$ $(\chi_m)$} \\ \hline
 Run1 & Run2 & Run1 & Run2 & Run1 & Run2 & Run1 & Run2 \\ \hline
$3.24 \pm 0.05$ &  $3.15 \pm 0.03$ & $2.13 \pm 2.48$ & $2.37 \pm 0.69$ & $1.51 \pm 0.14$ & $1.35 \pm 0.1$ & $2.78 \pm 1.27$ & $3.17 \pm 0.48$ \\ \hline
\end{tabular}
\vspace{-10pt}
\end{table*}

\begin{figure}[b!]
\vspace{-20pt}
\includegraphics*[width=\columnwidth]{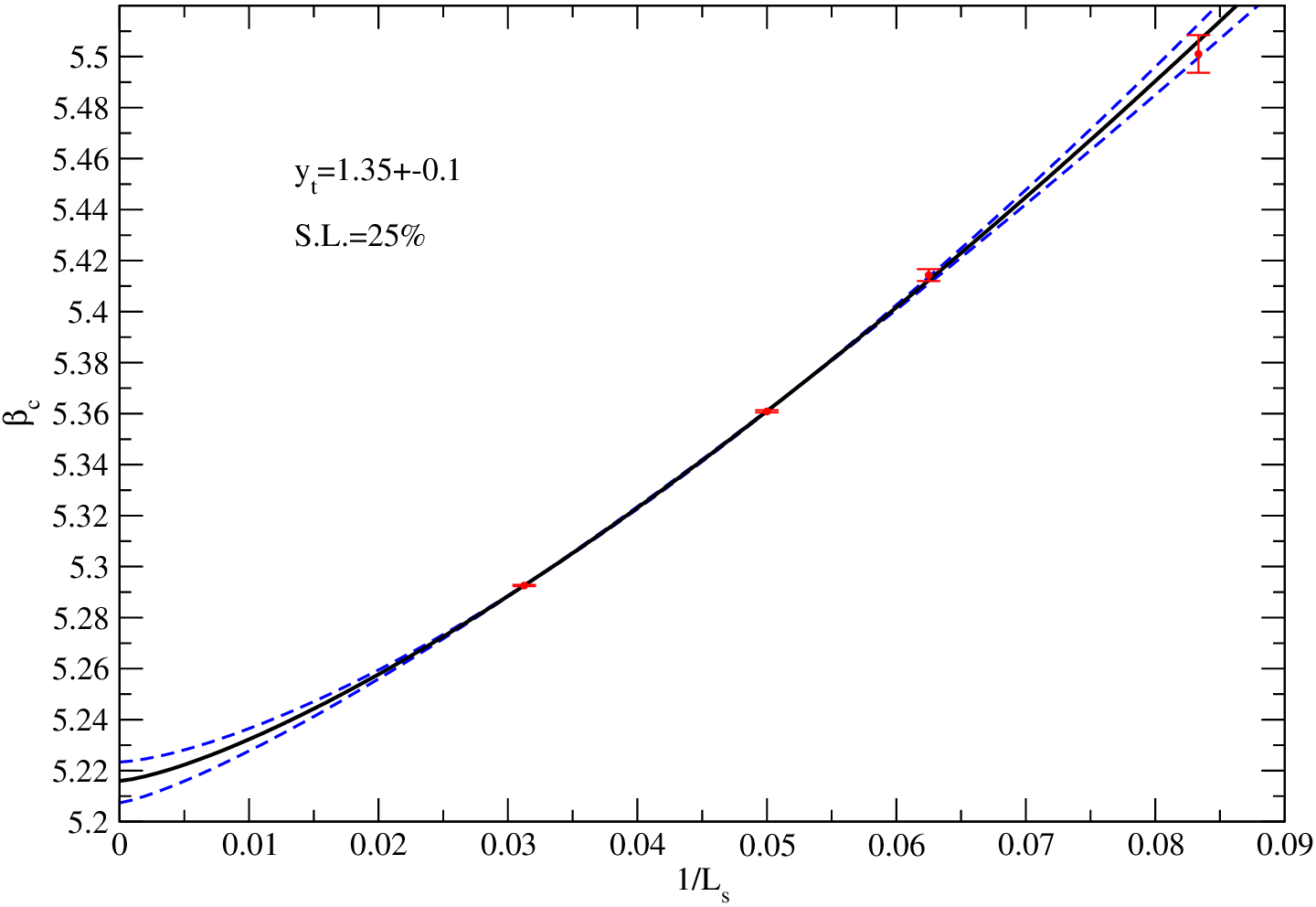}
\vspace{-45pt}
\caption{Pseudocritical coupling vs 1/$L_s$}\label{bc} 
\vspace{-20pt}
\end{figure}

The behavior of the $\rho$ susceptibility is shown in Fig.\,\ref{rho}. For $T<T_c$ $\rho$ is size indipendent showing that $\left<\mu\right>\neq 0$; for $T>T_c$ $\rho$ shows a growing linear with $L$ showing that $\left<\mu\right>$ is strictly zero in the limit $L\rightarrow \infty$. 

\begin{figure}[b!]
\vspace{-20pt}
\includegraphics*[width=\columnwidth]{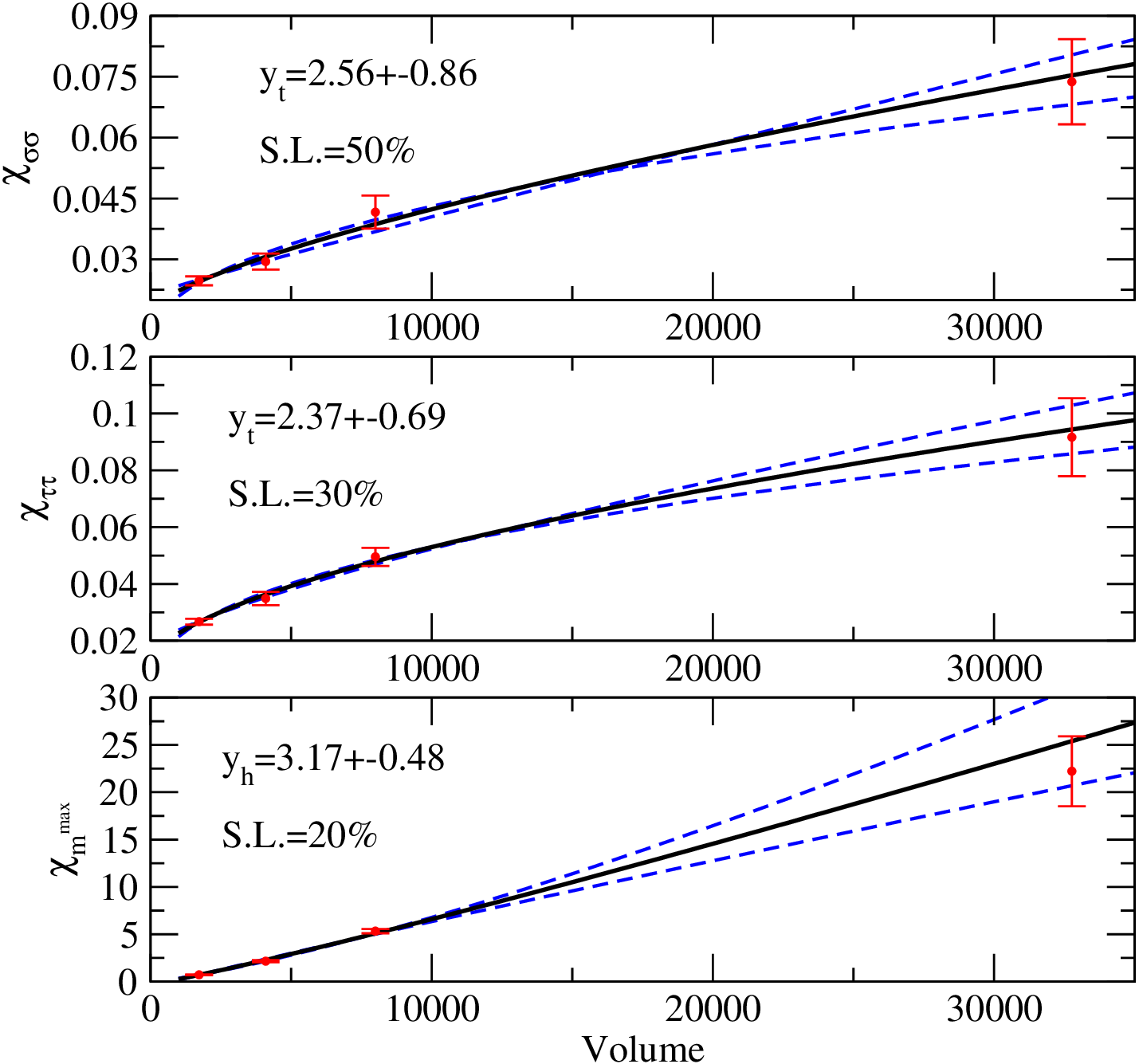}
\vspace{-45pt}
\caption{External lines show the 1-$\sigma$ confidence region for $y_h$}\label{chips} 
\vspace{-25pt}
\end{figure}

A power law fit of the $\rho$ peak as a function of $L$ gives a value for $y_t$ (Eq.~\ref{rhosca}). Other estimates for this exponent can be obtained from analogous fits (Fig.~\ref{bc}-\ref{chips}) for the pseudocritical coupling (Eq.~\ref{bcsca}) and the specific heat through $\chi_{ij}$, $\chi_{ee}$ and $\chi_{ie}$ (Eq.~\ref{cvsca}) as shown in Fig.~\ref{bc}-\ref{chips}. 
Result for the critical exponents are summarized in Table~\ref{exp}. These estimates tends to be greater than the expected values for $O(4)$ ($y_t=1.34$, $y_h=2.49$), $O(2)$ ($y_t=1.49$, $y_h=2.48$) and mean field ($y_t=3/2$, $y_h=9/4$). Moreover the exponent from $\rho$ seems to indicate a first order phase transition.
As a consistency check one can fit the peak of $\chi_m$ to read back from data the assumed magnetic exponent $y_h$ (Fig.~\ref{chips}). Also in this case we note a deviation from the starting value $y_h=2.49$.

We tested the $O(4)$ scaling behavior and found it inconsistent with our data. A first order transition cannot be excluded but our initial assumption for $y_h$ flaws the analysis. Moreover the staggered fermion action at $N_t=4$ is probably plagued by large lattice artifacts that suppress the critical behavior.

\end{document}